\documentclass[preprint,amsmath,amssymb,aps,pra]{revtex4-1}
\usepackage[utf8]{inputenc}
\usepackage{graphicx}
\usepackage{dcolumn}
\usepackage{mathrsfs,amsmath}
\usepackage{bm}
\usepackage{graphicx}
\usepackage{dcolumn}
\usepackage{bm}
\usepackage[mathlines]{lineno}
\modulolinenumbers[5]

\begin{document}

\preprint{}

\title[Zero-Point Motion of $p$-H$_2$]
{
Zero-Point Motion of Liquid and Solid Hydrogen}

\author{T. R. Prisk}
 \altaffiliation[Current Affiliation: ]{Lawrence Livermore National Laboratory, Livermore CA 94550, United States}
\author{R. T. Azuah}
 \altaffiliation{Department of Materials Science and Engineering, University of Maryland - College Park, MD 20742-2115, United States}
 \affiliation{Center for Neutron Research, National Institute of Standards and Technology, Gaithersburg MD, 20899-6100, United States}

\author{D. L. Abernathy}
\author{G. E. Granroth}
\author{T. E. Sherline}
\affiliation{Neutron Scattering Division, Oak Ridge National Laboratory, Oak Ridge, TN, 37831, United States}

\author{P. E. Sokol}
\affiliation{Department of Physics, Indiana University, Bloomington, IN, 47408, United States}

\author{J. Hu}
\author{M. Boninsegni}
\affiliation{Department of Physics, University of Alberta, Edmonton, AB, T6G 2E1, Canada}

\date{\today}

\begin{abstract}
We present an inelastic neutron scattering study of liquid and solid hydrogen carried out using the wide Angular Range Chopper Spectrometer at Oak Ridge National Laboratory.  From the observed dynamic structure factor, we obtained empirical estimates of the molecular mean-squared displacement and average translational kinetic energy.  We find that the former quantity increases with temperature, indicating that a combination of thermal and quantum effects are important near the liquid-solid phase transition, contrary to previous measurements.  We also find that the kinetic energy drops dramatically upon melting of the crystals, a consequence of the large increase in molar volume together with the Heisenberg indeterminacy principle.  Our results are compared with quantum Monte Carlo simulations based upon different model potentials.  In general, there is good agreement between our findings and theoretical predictions based upon the Silvera-Goldman and Buck potentials.
\end{abstract}

\maketitle

\section{Introduction}
The condensed phases of molecular hydrogen are systems of fundamental interest to quantum many-body physics.  Due to their light mass, the zero-point motion of hydrogen molecules makes a significant contribution to the atomic-scale structure and dynamics of liquid and solid hydrogen.  The importance of quantum-mechanical effects places the condensed phases of molecular hydrogen in a position between classical substances, on the one hand, and highly degenerate quantum fluids and solids, on the other\cite{Silvera}.  Although the condensed phases of molecular hydrogen are not believed to display superfluidity \cite{boninsegni2018}, even in reduced dimensions\cite{boninsegni2004,boninsegni2013},  quantum effects can be detected in the momentum distribution of liquid parahydrogen near melting \cite{Gernoth2007,Boninsegni2009}. Moreover, superfluidity has been predicted to occur at low temperature (on the order of 1 K) in small clusters of parahydrogen, comprising tens of molecules \cite{Sindzingre1991,Mezzacapo2006,Mezzacapo2007,Mezzacapo2008, Review_H2superfluid}. Besides superfluidity, current scientific interest in hydrogen encompasses a broad range of topics, including hydrogen storage materials \cite{Tozzini2013,Review_H2sorption}, planetary science \cite{Review_DenseHydrogen}, and thermonuclear fusion \cite{Souers,Review_H2fusion,NIF_Lawson}.  More generally, nuclear quantum effects are significant in materials comprised of light atoms \cite{Review_PIMCsolids} and in hydrogen-bonding substances \cite{Review_Water,Review_Guo}.

Despite its basic role in condensed hydrogen, the molecular momentum distribution is not fully understood at present.  Several groups have carried out inelastic neutron scattering measurements of the average molecular kinetic energy $\langle E_K\rangle$ of liquid \textit{para}-hydrogen \cite{Langel,CelliKE,ZoppiKE,Dawidowski,Andreani1}.  Unfortunately, experiments performed with the TOSCA \cite{CelliKE,ZoppiKE} and MARI \cite{Dawidowski} spectrometers yield conflicting values for average kinetic energy in the liquid state.  In particular, at 16.5 K, the former groups finds that $\langle E_K\rangle$ is 60.3(6) K, whereas the latter group obtains a value of 67.8(3) K.  Recent quantum Monte Carlo simulations of liquid \textit{para}-hydrogen, built upon different model pair potentials, yield similarly conflicting predictions for $\langle E_K\rangle$ \cite{H2_potentials}.  The discrepancy between the empirical estimates of $\langle E_K\rangle$ makes the choice between available microscopic models of the liquid an ambiguous one.  This situation sharply contrasts with $^4$He, where the momentum distribution can be predicted to a high degree of accuracy \cite{Prisk,Azuah1997,Diallo2,Nakayama}.

Another area of significant disagreement between inelastic neutron scattering experiments and quantum Monte Carlo simulations concerns the relative importance of quantum and thermal effects in the liquid-solid phase transition of hydrogen \cite{FA,Dusseault_MSD}.  Fernandez-Alonso \textit{et al.} examined the lowest rotational transition of solid hydrogen by means of the IN20 spectrometer.  They obtained a temperature independent value for the molecular mean-squared displacement $\langle u^2\rangle$ of 0.56 $\textrm{\AA}^{2}$.  On that basis, they concluded that thermal effects play a negligible role in the hcp solid and that its properties are wholly dominated by quantum-mechanical effects.  In contrast, quantum Monte Carlo calculations predict that $\langle u^2\rangle$ increases from $0.519(2) \textrm{ \AA}^2$ to $0.622(2) \textrm{ \AA}^2$ as the temperature is increased from 4 K to 13.8 K.  This suggests that a combination of quantum and thermal effects is at play near the liquid-solid phase transition.

In this paper, we present an inelastic neutron scattering study of liquid and solid hydrogen under saturated vapor pressure.  In particular, we report high-precision measurements of the molecular mean-squared displacement and average translational kinetic energy.  The experiment was performed using the wide Angular Range Chopper Spectrometer at Oak Ridge National Laboratory.  Empirical estimates of the molecular mean-squared displacement were obtained from the wavevector dependence of the first rotational state transition.  The momentum distribution of the hydrogen molecules was inferred from recoil scattering \textit{via} the Impulse Approximation.  As shown in detail below, we find good agreement between the new observations and quantum Monte Carlo simulations based upon the Silvera-Goldman and Buck potentials.

For convenience, we introduce the following notation to refer to transitions between the rotational states of molecular hydrogen.  We write $(J, J^\prime)$ to refer to a transition where $J$ and $J^\prime$ are the initial and final rotational quantum numbers, respectively.  Throughout, the hydrogen molecules remain in their electronic and vibrational ground states.

\section{Experimental Approach}

\subsection{Wide Angular Range Chopper Spectrometer}
We carried out an inelastic neutron scattering study of liquid and solid hydrogen under saturated vapor pressure using the wide Angular Range Chopper Spectrometer (ARCS) at the Spallation Neutron Source\cite{ARCS,Stone}.  This instrument is a direct geometry, time-of-flight spectrometer.  Incident neutron energies between 15 meV and 5000 meV are available from the decoupled poisoned water moderator.  A $T_0$ chopper blocks prompt radiation released by the target during spallation.  The incident neutron energy is chosen \textit{via} time-of-flight by a Fermi chopper located before the sample.  The secondary spectrometer consists of a cylindrical bank of 920 position sensitive $^3$He detectors spanning -28$^\circ$ to +135$^\circ$ in the horizontal plane.  There are two low efficiency beam monitors, one located after the Fermi chopper and another located just before the beam stop. The beam profile observed at these monitors is used to determine the incident neutron energy $E_i$ and moderator emission time.

Measurements were carried out at each experimental condition using 30 meV ($\lambda = 1.65 \textrm{ \AA}^{-1}$) and 500 meV ($\lambda = 0.404 \textrm{ \AA}^{-1}$) incident neutrons.  The 30 meV data sets reported in this paper were acquired with either 834 $\mu\textrm{A}\cdot\textrm{hrs}$ or 1668 $\mu\textrm{A}\cdot\textrm{hrs}$ of proton charge delivered to the target, whereas the 500 meV data sets correspond to 3330 $\mu\textrm{A}\cdot\textrm{hrs}$.  For the $E_i = 30 \textrm{ meV}$ measurements, we ran a Fermi chopper that nominally possessed 1.52 mm slit thickness, 0.35 slat thickness, a 50 mm radius, and a blade curvature of 0.580 m.  This Fermi chopper was set to a frequency of 300 Hz, and the $T_0$ chopper was operated at 90 Hz.  For the $E_i = 500 \textrm{ meV}$ measurements, we employed a Fermi chopper with  nominal 0.51 mm slit thickness, 0.35 mm slat thickness, a 50 mm radius, and a blade curvature of 1.535 m.  However, as the slit package for the 500 meV measurements is rather tight, manufacturing uncertainties imply the effective slit thickness is finer than designed.  Ray-tracing Monte Carlo simulations, discussed further in the next section, suggest that the effective slit width is 0.192 mm.  This chopper frequency was set to 480 Hz, while the $T_0$ frequency was set to 120 Hz.  The chopper frequencies were chosen to maximize the incident neutron flux at the desired incident energies.

The sample environment chosen for this experiment was a closed-cycle refrigerator with aluminum tails.  Research-grade hydrogen was loaded \textit{in situ} to the sample cell from a gas handling system.  This system includes a refrigerated vessel containing a chromium oxide catalyst, allowing cooled gas to undergo \textit{ortho} to \textit{para} conversion before being loaded.  We employed an aluminum plate cell that was oriented at thirty degrees relative to the incident beam.  The sample space was 71 mm wide, 50 mm tall, and 0.508 mm deep.  A pocket below the sample space contained a Cr(II) oxide catalyst\cite{Catalyst}, and this catalyst was in continuous contact with the condensed hydrogen during the experiment.  The temperature of the condensed hydrogen sample was inferred from the observed vapor pressure using the expression given by Souers \textit{et al}\cite{Souers_P}.

The history of the condensed hydrogen sample is as follows.  Immediately after loading hydrogen to the cell, we cooled the sample to 5.0 K and conducted neutron scattering measurements.  We found that the initial mole fraction of \textit{para}-hydrogen within the sample was 89.41(6)\%.  We subsequently melted the hydrogen sample and allowed it to equilibrate with the chromium oxide catalyst contained in the sample cell for approximately thirteen hours.  The resulting \textit{para}-hydrogen concentration was 99.70(2)\%.  These concentrations were inferred from the relative intensities of the $(0, 1)$ and $(1, 0)$ peaks, as described below.  Measurements were then carried out in the following order: 12.7 K, 10.0 K, 8.4 K, 5.0 K, and 16.5 K.  Both incident energies were employed before changing temperature.  The scattering from the empty cell and sample environment was measured at 15 K.

The double-differential scattering cross section was transformed to the dynamic structure factor by means of Mantid\cite{Mantid} and the Data Analysis and Visualization Environment\cite{DAVE}.

\subsection{Instrumental Resolution}
In order to obtain accurate empirical estimates of $\langle E_K\rangle$, it is necessary to account for the effects of instrumental resolution upon the observed dynamic structure factor.  At a spallation neutron source, the resolution function of a Fermi chopper spectrometer is determined by the velocity-time distribution of the source and the response functions of the various instrument components, and consequently it may assume an asymmetric form\cite{Carpenter}.  In this case, the observed peaks are significantly broader than the instrumental resolution function, making the detailed lineshape of the latter unimportant for present purposes.  

Therefore, for the $E_i = 500 \textrm{ meV}$ measurements, we adopt a Gaussian approximation, according to which the moderator pulse width, Fermi chopper pulse width, and detector time uncertainty combine in quadrature to determine the energy resolution \cite{Windsor,ResolutionCalc}.  We obtained a moderator pulse width of 3.095 $\mu$s from Monte Carlo N-Particle Transport Code System (MCNPX) simulations of the decoupled water moderator \cite{Iverson}.  The observed profile width in the first beam monitor, namely 3.19 $\mu$s, was taken as an estimate of chopper pulse width.  The detector time uncertainty is given by the width of a detector divided by the neutron final velocity.  The calculated resolution width decreases from 15.5 meV at $E = 0$ meV to 4.8 meV at $E = 400$ meV.

We performed a ray-tracing Monte Carlo simulation of the ARCS instrument with the McStas software suite\cite{McStas1,McStas2,ARCS_sim}.  For $E_i = 30 \textrm{ meV}$, we found excellent agreement between the simulated and observed monitor spectra with no modification of the instrument parameters from their nominal values.  We furthermore found the simulation correctly reproduced the elastic resolution function determined by measurements of a vanadium plate.  For $E_i = 500 \textrm{ meV}$, excellent agreement between simulated and observed monitor spectra was found after refining the value of the effective Fermi chopper slit width.  Because the primary spectrometer functions analogously to a pinhole camera, where the Fermi chopper acts as the pinhole, the second monitor is especially sensitive to the description of the moderator. Thus, the outcome of the simulations confirms that the MCNPX description of the moderator is valid, and that the moderator pulse width used in our resolution calculations is correct.

\subsection{Multiple Scattering}
The sample geometry was chosen to minimize the amount of multiple scattering.  Ideally, one would like each neutron to interact once with the sample before reaching the detector.  However, in practice, neutrons can undergo several scattering events within the sample, and, for these neutrons, the simple relationship between the double-differential scattering cross section and the dynamic structure factor is lost.

Multiple scattering is expected to be negligible here given the macroscopic scattering cross section of the condensed hydrogen and the geometry of the sample cell.  The total neutron scattering cross section of liquid hydrogen at $E_i = 500 \textrm{ meV}$ is approximately 44 barns/molecule \cite{H2barns,Seiffert,Grammer}.  At 16.5 K, the number density of liquid hydrogen is 0.0223 $\textrm{ \AA}^{-3}$ \cite{Souers}.  Accordingly, the macroscopic scattering cross section is 0.981 $\textrm{ cm}^{-1}$ and the neutron mean free path is 1.019 cm.  Given that the plate cell had a thickness of 0.508 mm and was oriented at 30$^\circ$ relative to the incident beam, the fraction of scattered neutrons is approximately 5.6\%.

\section{Results}

\subsection{Dynamic Structure Factor}

\begin{figure}[h]
\centering
  \includegraphics[width=0.9\textwidth,height=0.9\textheight,keepaspectratio]{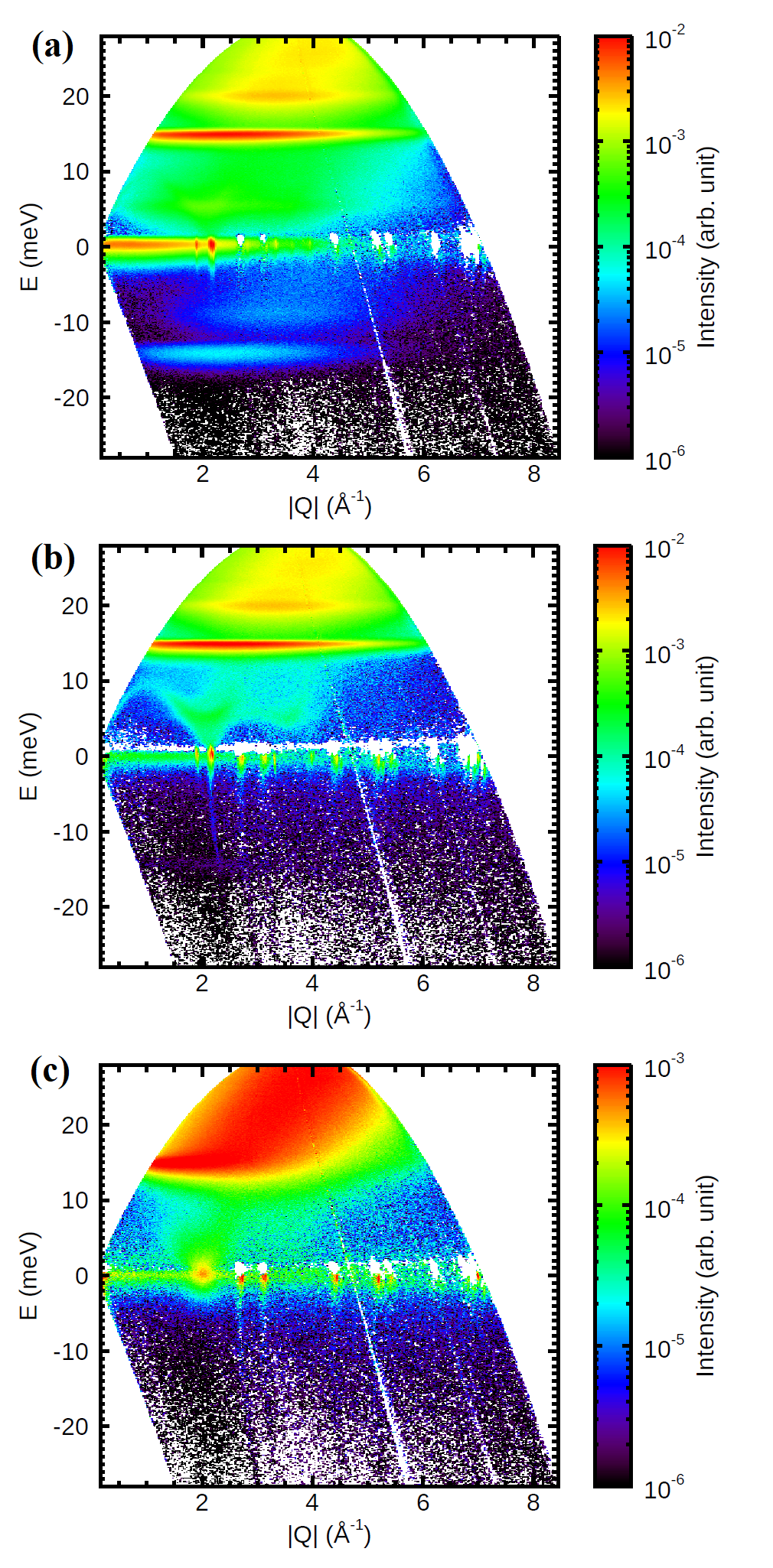}
  \caption{The dynamic structure factor of solid and liquid hydrogen obtained with a 30 meV incident neutron energy: (a) the solid phase with $X_{\textrm{para}} = 89.41(6)\%$ at $T = 5 \textrm{ K}$; (b) the solid phase with $X_{\textrm{para}} = 99.70(2)\%$; and (c) the liquid phase at 16.5 K.}
  \label{fgr:30meV}
\end{figure}

We first consider the dynamic structure factor obtained with a 30 meV incident neutron energy.  Figure \ref{fgr:30meV} (a) illustrates $S(Q, E)$ of the initial solid hydrogen sample.  Along the elastic line ($E = 0$), one observes elastic incoherent scattering from \textit{ortho}-hydrogen as well as Bragg reflections of the hcp crystal.  We did not employ a radial collimator in this experiment, and so there is imperfect subtraction of the background signal originating from the aluminum tails of the closed-cycle refrigerator.  Between 0 meV and +15 meV, one observes the phonon density of states, which peaks near +5 meV\cite{H2_DOS}.  The $(0, 1)$ transition appears as a sharp peak near +15 meV.  Beyond the rotational transition, there are combinations of this transition with lattice vibrations, and these exhibit a local maximum near +20 meV.  At -15 meV, one sees upscattering due to the $(1, 0)$ rotational transition.  Lastly, combinations of that rotational transition with lattice vibrations peak near -10 meV. 

In Figure \ref{fgr:30meV} (b), we display the dynamic structure of the hydrogen sample after equilibration with the catalyst contained in the sample cell.  The  upscattering signal and elastic incoherent scattering have nearly, though not completely, disappeared.  Along the elastic line, the (110), (101), (110), (201), and (004) Bragg reflections are clearly seen.  The (002), (102), (200), and (112) peaks are expected to have low intensity, and they are not found.  The (103) peak is expected to be observed, but it appears to be obscured by the background.  We cannot judge whether the sample is polycrystalline or a ``true'' powder on the basis of the diffraction pattern.  In panel (a), the observed signal from the phonon density of states consists of incoherent inelastic scattering from \textit{ortho}-hydrogen and coherent inelastic scattering from \textit{para}-hydrogen.  In panel (b), the signal originates from coherent inelastic scattering from \textit{para}-hydrogen alone.

The dynamic structure factor of the liquid at 16.5 K is shown in Figure \ref{fgr:30meV}.  Here one observes coherent quasi-elastic scattering and the collective excitations of the liquid state\cite{liquidH2_phonons1,liquidH2_phonons3}.  The $(0, 1)$ transition is no longer sharply defined, but instead blends smoothly and continuously with the multi-phonon spectrum.  It is apparently broadened by translational diffusion, which has a quasi-elastic width on the order of a few meV\cite{nH2_diffusion}.  For comparison, we note that similar features are found in the dynamic structure factor of \textit{normal}-hydrogen within porous media, though with the modification that the molecules adsorbed to the pore walls are hindered, rather than free, rotors\cite{H2_MCM41}.

\begin{figure}[h]
\centering
  \includegraphics[width=0.8\textwidth,height=0.8\textheight,keepaspectratio]{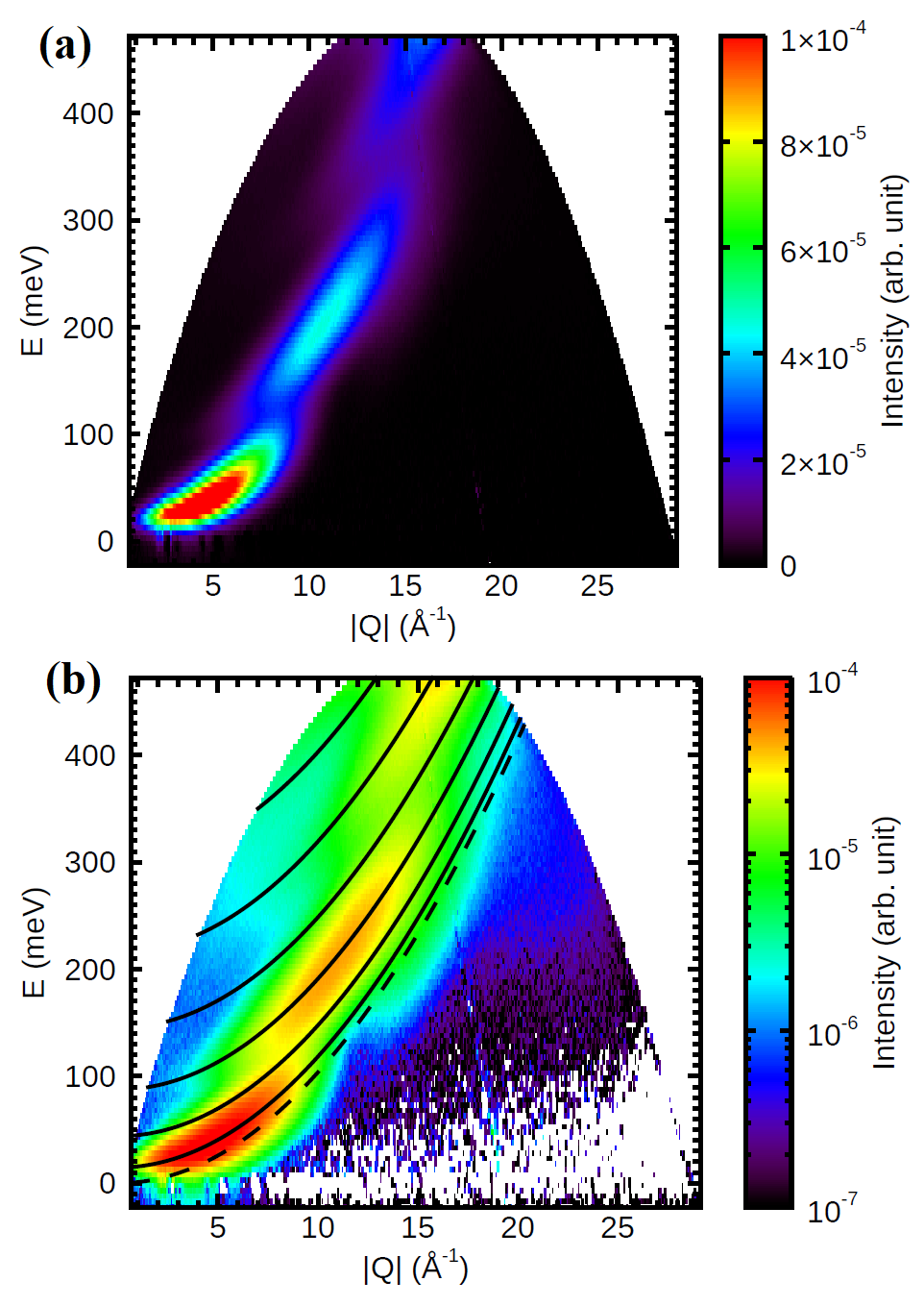}
  \caption{The dynamic structure factor of solid hydrogen obtained with a 500 meV incident neutron energy.  In panel (a), $S(Q, E)$ is shown on a linear intensity scale for $X_{\textrm{para}} = 99.70(2)\%$ at $T = 5 \textrm{ K}$.  In panel (b), the same data is shown on a logarithmic intensity scale.  The dashed line indicates the molecular recoil dispersion, and the solid lines indicate the dispersions of the $(0, 1)$, $(0, 2)$, $(0, 3)$, $(0, 4)$, $(0, 5)$, and $(0, 6)$ transitions.}
  \label{fgr:500meV}
\end{figure}

We now turn to measurements obtained with 500 meV incident neutrons.  Figure \ref{fgr:500meV} (a) displays the dynamic structure factor of solid hydrogen at 5 K.  Panel (b) shows the same data on a logarithmic intensity scale, where the superimposed dashed line represents free molecular recoil: $E_R = \hbar^2Q^2/2m$.  The recoil line is split by the internal rotational transitions of the molecule.  Most prominent in the spectrum are the $(0, 1)$, $(0, 3)$, $(0, 5)$ transitions.

\subsection{Ortho-Para Concentrations}
\begin{figure}[h]
\centering
  \includegraphics[width=0.8\textwidth,height=0.8\textheight,keepaspectratio]{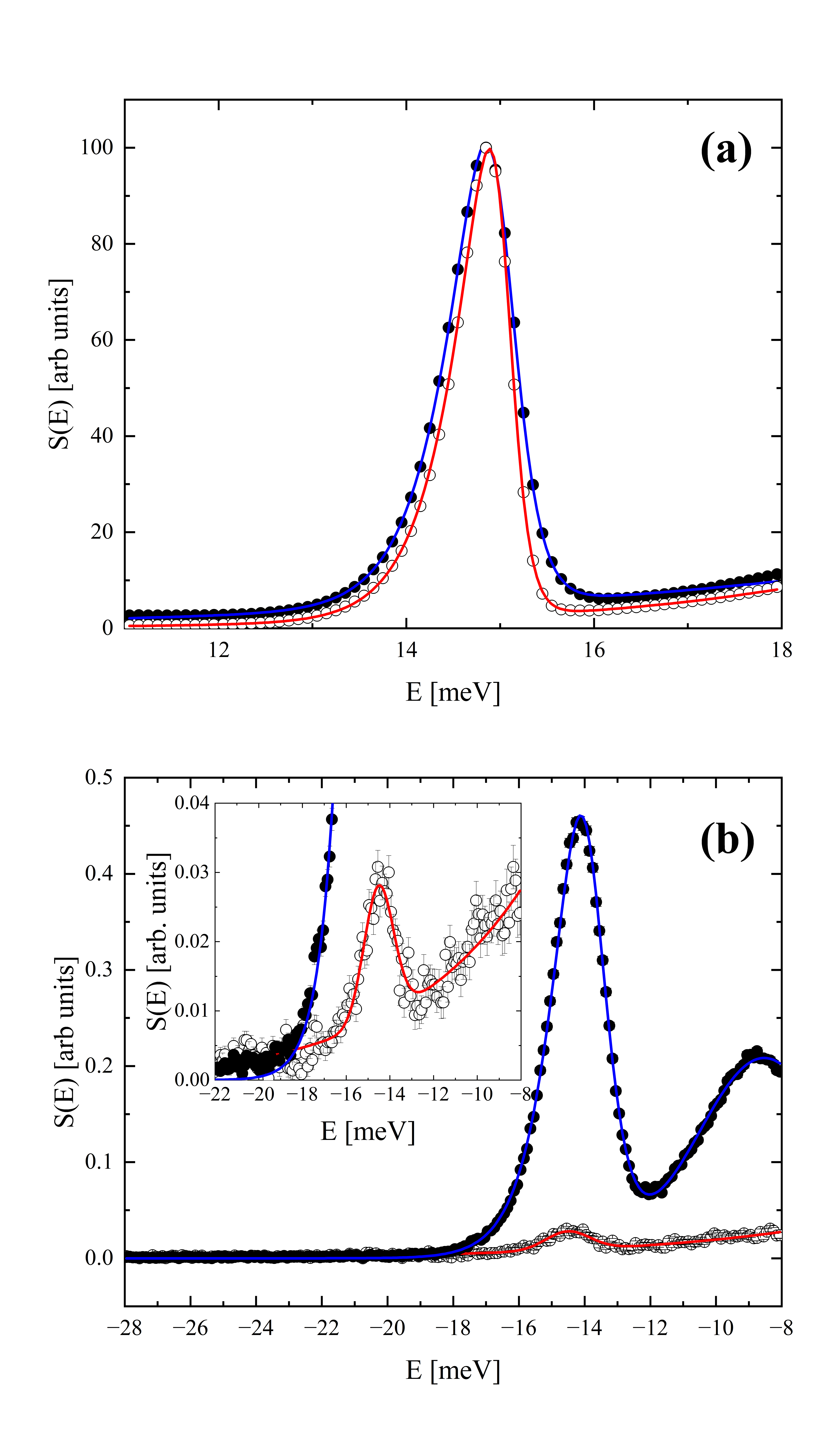}
 \caption{The inelastic scattering function $S(E)$ of solid hydrogen near the (a) $(0, 1)$ transition and the (b) $(1, 0)$ transition.  Data from the initial (final) sample is shown as closed (open) circles.  Fits, described in the text, are shown as solid curves.  The data in both panels has been scaled so that the maxima in panel (a) occur at 100 arbitrary units.  Where not visible, error bars are smaller than symbol sizes.}
  \label{fgr:ortho}
\end{figure}

The concentration $X_p$ of \textit{para}-hydrogen within the sample may be inferred from the relative intensities of the $(0, 1)$ and $(1, 0)$ transitions.  To first order, the integrated intensities of these peaks are given by the following expressions\cite{Sears1,Sears2,YK}:
\begin{align*}
A_{01} &= N_p \cdot 3\sigma_i j_1^2(Qa)e^{-2W_p}\\ 
A_{10} &= N_o \cdot \frac{1}{3}\sigma_i j_1^2(Qa)e^{-2W_o}
\end{align*}
Here $N_p$ and $N_o$ are the number of \textit{para}-hydrogen and \textit{ortho}-hydrogen molecules; $\sigma_i$ is the incoherent scattering cross section of atomic hydrogen; $j_n$ is a spherical Bessel function of order \textit{n}; $a =  0.3707\textrm{ \AA}$ is the radius of gyration of the hydrogen molecule; and $e^{-2W_p}$ and $e^{-2W_o}$ are the Debye-Waller factors of \textit{para}- and \textit{ortho}- hydrogen.  If one assumes that the Debye-Waller factors for the two species are identical, then it follows that the concentration of \textit{para}-hydrogen is the following:
\begin{equation}
    X_p = \frac{N_p}{N_p + N_o} = \frac{A_{01}}{A_{01} + 9A_{10}}
\end{equation}
To extract the peak intensities, $A_{01}$ and $A_{01}$, we first integrated the dynamic structure factor over $Q$ to obtain the inelastic scattering function $S(E)$.  Figure \ref{fgr:ortho} compares the observed $S(E)$ of the sample immediately after condensation and of the sample after equilibration with the catalyst.  To obtain the integrated intensities of the $(0, 1)$ and $(1,0)$ peaks, we represented them by an asymmetric double sigmoidal function $f(E)$, and the remaining scattering by a Gaussian:
\begin{equation}\label{eq:asym2sig}
f(E) = f_S\cdot \frac{1}{1 + e^{-(E - E_c)/w}}\left(1 - \frac{1}{1 + e^{-(E - E_c)/w^\prime}} \right)   
\end{equation}
Here $f_S$ is a scale factor, $E_C$ is the ``center'' of the peak, and $w$ and $w^\prime$ are constants that control the shape of the peak.  After obtaining this parameterized description of the peak, we obtained their intensities by numerical integration, as we have not found a closed, analytic expression for an integral over this peak shape. 

\section{Discussion}

\subsection{Empirical Estimates of $\langle u^2\rangle$}
We now consider the molecular mean-squared displacement of our solid hydrogen sample.  As discussed in the preceding section, the integrated intensity of the first rotational transition is proportional to the product of a rotational form factor and a Debye-Waller factor.  Here we retain the series expansion of the form factor up to fifth order:
\begin{equation}\label{eq:formfactor}
    \begin{split}
        A_{01}(Q) = S^\prime\cdot\left(3j_1^2(Qa) + 7j_3^2(Qa) + 11j_5^2(Qa)\right)\exp\left[-\frac{Q^2\langle u^2\rangle}{3}\left(1 - \alpha Q^2\right)\right]
    \end{split}
\end{equation}
$S^\prime$ is an overall scale factor and $\alpha$ is an anharmonic coefficient.  When fitting the integrated intensity versus $Q$, the adjustable parameters are: $S^\prime$, $\langle u^2\rangle$, and $\alpha$.

\begin{figure}[h]
\centering
  \includegraphics[width=0.9\textwidth,height=0.9\textheight,keepaspectratio]{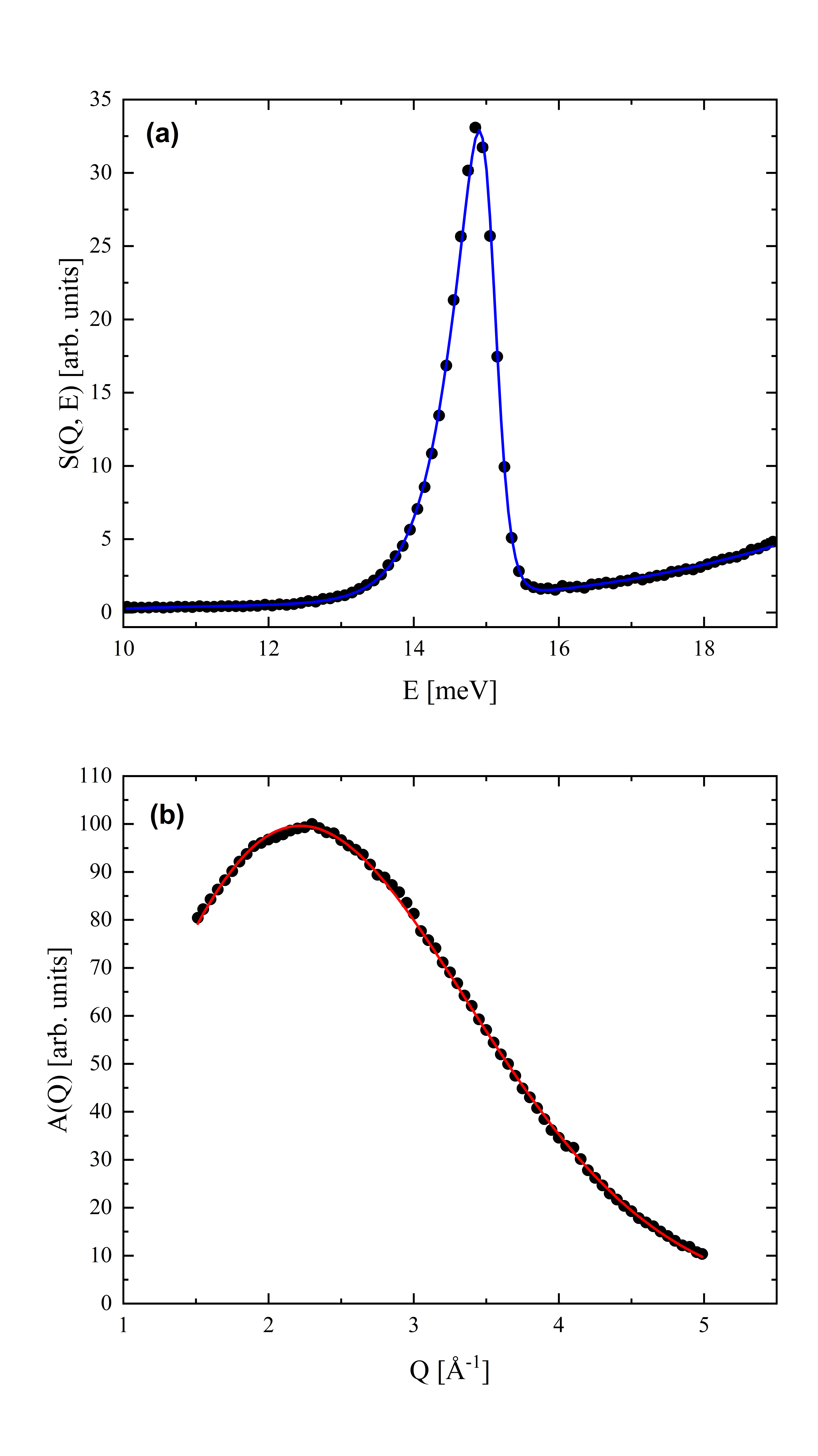}
 \caption{(a) The dynamic structure factor $S(Q, E)$ of solid hydrogen at $Q = 2.5 \textrm{ \AA}^{-1}$ and $T = 12.7 \textrm{ K}$.  (b) The integrated intensity of the first rotational transition at the same temperature as a function of wavevector transfer.}
  \label{fgr:MSDfit}
\end{figure}

The values of $\langle u^2\rangle$ and $\alpha$ in the solid phase were obtained from the first rotational transition as follows.  Figure \ref{fgr:MSDfit} (a) plots $S(Q, E)$ at $Q = 2.5 \textrm{ \AA}^{-1}$ and $T = 12.7 \textrm{ K}$.  The scattering has been fit to the asymmetric double sigmoidal function, given in Equation \ref{eq:asym2sig}, and a cubic polynomial.  Figure \ref{fgr:MSDfit} (b) plots $A(Q)$ as a function of $Q$ for this same temperature.  We first carried out a non-linear least-squares fit according to Equation \ref{eq:formfactor}.  Unfortunately, all three adjustable parameters are strongly correlated with one another.  We then employed the differential evolution algorithm\cite{DiffEvAlog} with a 5\% $\chi^2$-tolerance, and took the respective pointwise errors to represent the uncertainties on the adjustable parameters.

Table \ref{tbl:results} compiles our empirical estimates of $\langle u^2\rangle$ and $\alpha$.  We find that molecular mobility and anharmonicity grow with increasing temperature.  Moreover, both quantities are the same for the initial and equilibriated solid samples, at least within experimental precision.  No values could be obtained for the liquid phase because the $(0, 1)$ transition does not appear as a sharp peak, but instead merges continuously with the multiphonon spectrum.

\begin{figure}[h]
\centering
  \includegraphics[width=0.9\textwidth,height=0.9\textheight,keepaspectratio]{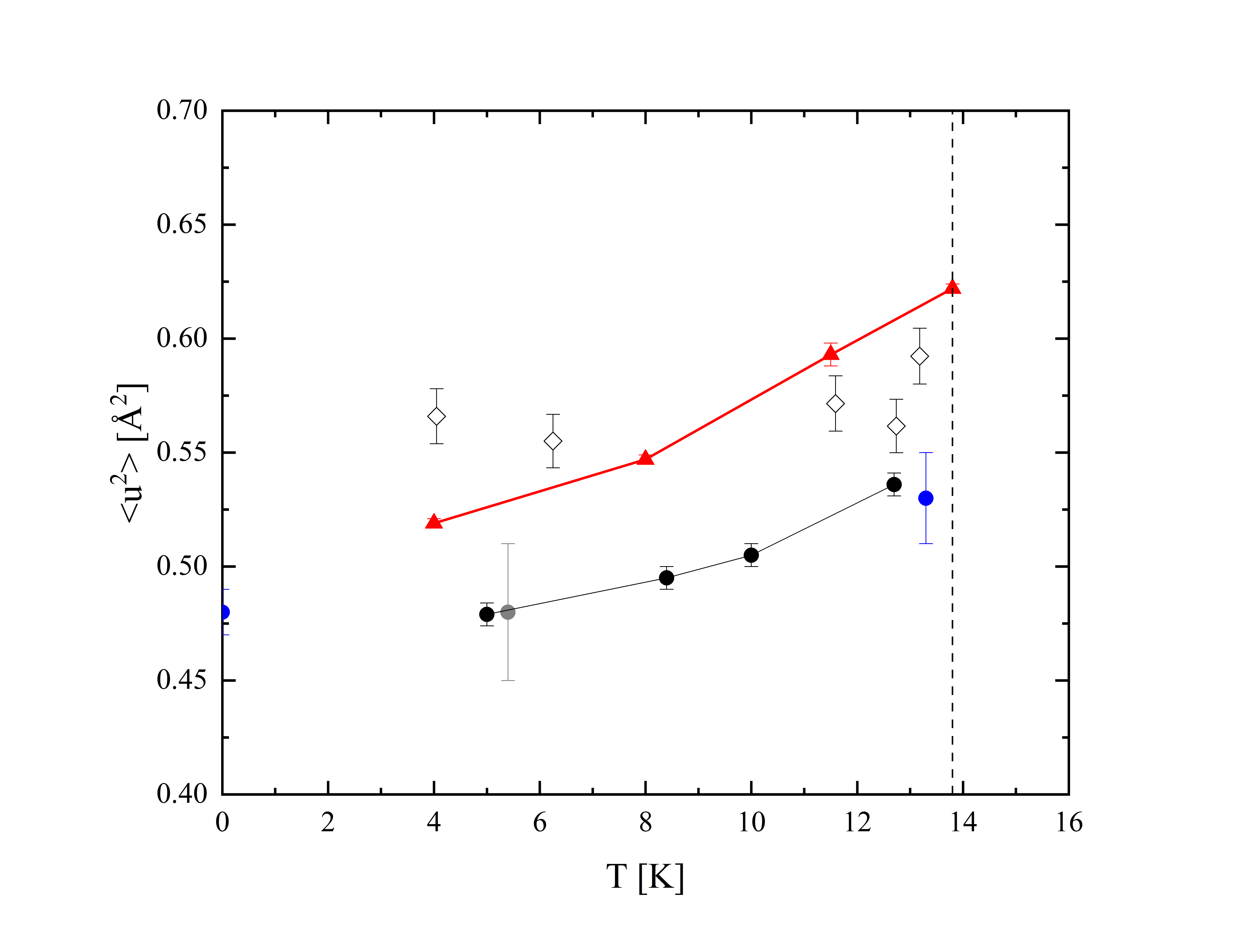}
 \caption{Theoretical and empirical estimates of the molecular mean-squared displacement of solid \textit{para}-hydrogen.  Symbol designations: ARCS result for $X_{para} = 99.70(2)\%$ (closed circle); triple-axis measurement\cite{Nielsen} (gray circle); phonon density of states\cite{H2_DOS} (blue circle); IN20 measurement\cite{FA} (open diamonds); quantum Monte Carlo simulation\cite{Dusseault_MSD} (red triangles).  The dashed vertical line indicates the liquid-solid phase transition at 13.8 K.  }
  \label{fgr:MSDsummary}
\end{figure}

Theoretical and experimental values for $\langle u^2\rangle$ of solid hydrogen under saturated vapor pressure are shown in Figure \ref{fgr:MSDsummary}.    The ARCS data set stands in excellent agreement with the previous triple-axis measurement of Nielsen\cite{Nielsen} and with values inferred from the phonon density of states\cite{H2_DOS}.  On the other hand, our results are at variance with the IN20 experiment\cite{FA}, where $\langle u^2\rangle$ was found to possess a temperature independent value of $0.56 \textrm{ \AA}^2$.  Quantum Monte Carlo simulations\cite{Dusseault_MSD} are in semi-quantitative agreement with experiment: the theory predicts the correct behavior with temperature, although there is an overall shift to larger values of $\langle u^2\rangle$.

We here define the Lindemann ratio $\delta = \sqrt{\langle u^2\rangle}/a$, where $a$ is the so-named lattice parameter of the hcp unit cell.  This quantity characterizes molecular mobility relative to the size of the crystal unit cell.  Employing the lattice parameters found by Krupskii \textit{et al}\cite{Krupskii_H2}, we find that $\delta$ increases from 0.183 to 0.193 as the temperature is increased from 5 K to 12.7 K.  This suggests that both thermal and quantum effects play roles in the liquid-solid phase transition of molecular hydrogen.

\subsection{Empirical Estimates of $\langle E_K\rangle$}
\begin{figure*}
 \centering
 \includegraphics[width=0.9\textwidth,height=0.9\textheight,keepaspectratio]{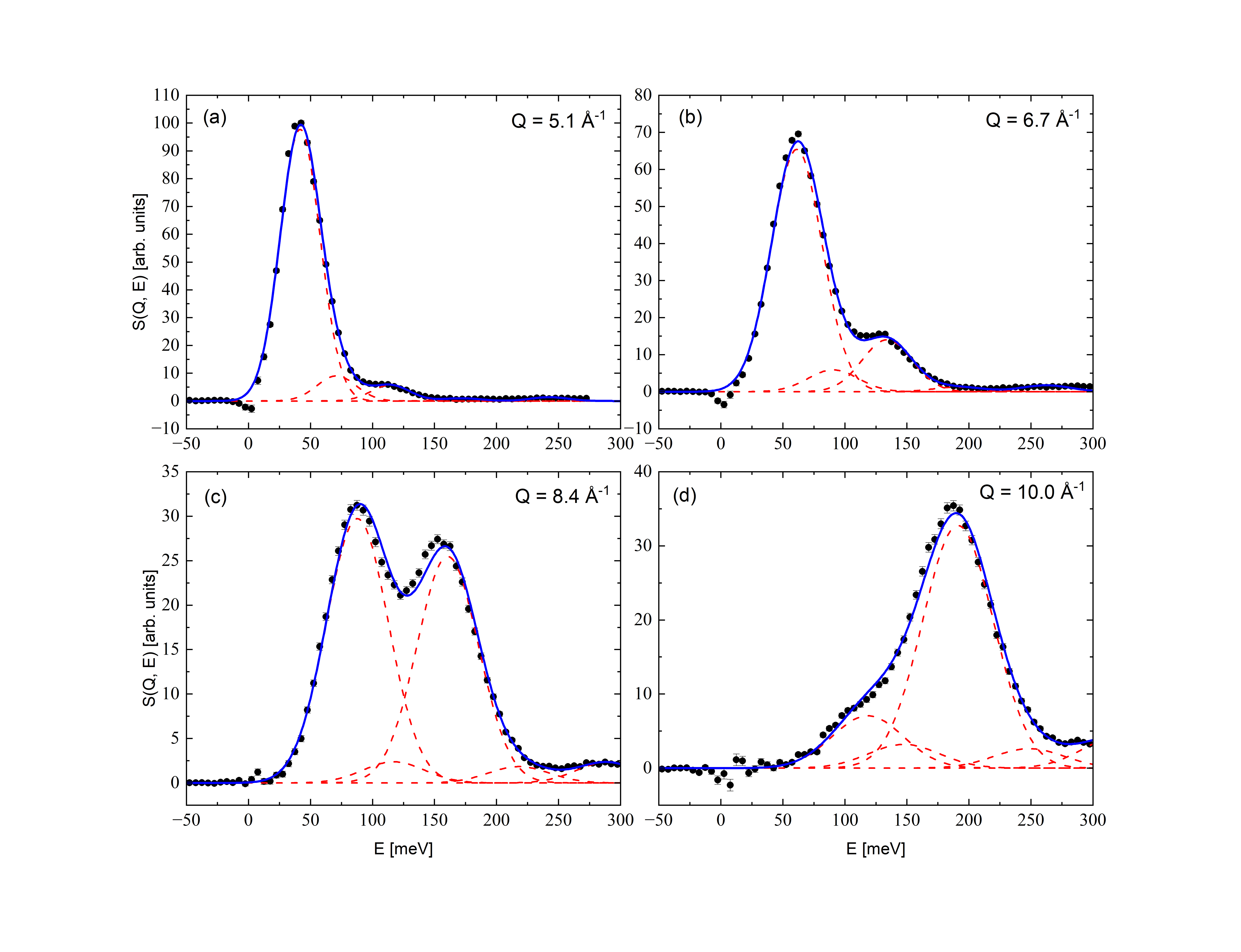}
 \caption{The dynamic structure factor of solid hydrogen at 12.7 K: experimental data (solid black points); model fit (solid blue curve); and resolution-broadened Gaussian components of the model (dashed red curves).  Error bars are due to Poisson counting statistics and they represent one standard deviation.}
 \label{fgr:cuts}
\end{figure*}

At high energies, the dynamic structure factor of liquid and solid hydrogen consists of the molecular recoil dispersion, though split by internal rotational transitions\cite{Langel}.  This can be seen in Figures \ref{fgr:500meV} and \ref{fgr:cuts}.  Ideally, one would like to determine the position, intensity, and lineshapes of the peaks contained in the spectrum wholly empirically.  Despite the quality of the data ($<2 \%$ statistical noise and $< 15 \textrm{ meV}$ energy resolution), this cannot be done, as the peaks are broad and overlapping.  Therefore, it is necessary to adopt \textit{a priori} assumptions in the data analysis, and our empirical estimates of the average molecular kinetic energy will be valid to the extent that these assumptions are valid. 

We make the following assumptions: (1) the molecular momentum distribution is Gaussian; (2) the incoherent approximation is valid; (3) the impulse approximation is valid; (4) the rotational transitions in the liquid and solid states occur at the same energies as their counterparts in the gaseous phase; and (5) the scattering from \textit{ortho}-hydrogen is negligible.  The first and last assumptions are adopted for simplicity and, ultimately, their justification turns on their adequacy in describing the observed scattering.  The second assumption is appropriate since the static structure factor is $S(Q) \approx 1$ for $Q \ge 5 \textrm{ \AA}^{-1}$\cite{CelliStructure}.  Our third assumption cannot be given a firm foundation, as there is currently no theory of final state effects in condensed hydrogen available.  We offer the qualitative argument that the asymmetrical broadening produced by final state interactions should be small when the momentum distribution is broad, as is the case in hydrogen.  For (4), we appeal to Raman spectroscopy measurements of condensed hydrogen\cite{Raman_pH2}.  In the liquid state, the energy of the $(0, 2)$ transition is reduced by 170 $\mu$eV from its value in the gaseous state.  In the solid, this transition is split into a triplet with a spacing of 250 $\mu$eV between the sub-levels.  These perturbations are far too small to observe \textit{via} ARCS.

On the basis of these assumptions, we suppose that the intrinsic $S(Q, E)$, at a given $Q$, consists of a series of peaks whose positions are shifted from the recoil energy by the relevant rotational transition energies, and whose lineshapes are determined by the momentum distribution of the molecules.  In particular, the dynamic structure factor is a sum of Gaussian peaks whose intrinsic widths are proportional to the average molecular kinetic energy:
\begin{equation}
    S(Q, E) =  \sum_{J = 1}^{5} \frac{A_J(Q)}{\sqrt{2\pi \sigma_J^2}}\exp\left[-\frac{E - E_J - E_R}{2\sigma_J^2}\right]
    \label{eq:model}
\end{equation}
$A_J(Q)$, $E_J$, and $\sigma_J^2$ are the integrated intensity, transition energy, and the observed second moment of the $J$$^{\textrm{th}}$ peak, respectively.  For $E_J$, we use the values reported in Ref\cite{Stoicheff}. The intrinsic width of the peak combines in quadrature with the inelastic energy resolution to yield the observed width: $\sigma_J^2 =\sigma^2 + \sigma_R^2$.  At a particular value of $Q$, all of the peaks in the spectrum share a common value of $\sigma$.  Thus, there are up to six adjustable parameters in the model $S(Q, E)$: the integrated intensities $A_J(Q)$ and the intrinsic second moment $\sigma^2$.  The average kinetic energy is: $\langle E_K\rangle = \left(3m/2\hbar^2\right)\left(\sigma/Q\right)^2$.

Figure \ref{fgr:cuts} plots representative fits to the scattering data at $T = 12.7 \textrm{ K}$.  In the solid phase, the dynamic structure factor was fit using Equation \ref{eq:model} at wavevectors within the range $5.0 \textrm{ \AA}^{-1} \le Q \le 10.0 \textrm{ \AA}^{-1}$ and energies within the range $-100 \textrm{ meV} \le E \le +300 \textrm{ meV}$.  In the liquid state, we modeled a narrower range, with $5.0 \textrm{ \AA}^{-1} \le Q \le 8.0 \textrm{ \AA}^{-1}$ and  $-100 \textrm{ meV} \le E \le +200 \textrm{ meV}$.  The model provides a good, though not perfect, description of the data.  For example, in the 12.7 K data set, typical values of $\chi^2$ fall between one and ten.  However, for $Q < 7 \textrm{ \AA}^{-1}$, $\chi^2$ reaches as high as twenty.  We attribute this to two distinct factors.  First, the number of neutron counts at low $Q$ is apparently sufficiently high that systematic effects (such as the definition of the energy scale, details of the peak lineshape, or the Gaussian approximation of the resolution function) begin to impose a statistical penalty.  Second, the scattering above 200 meV is relatively flat, and the model does not fully capture this aspect of the data.

\begin{figure}[h]
\centering
  \includegraphics[width=0.9\textwidth,height=0.9\textheight,keepaspectratio]{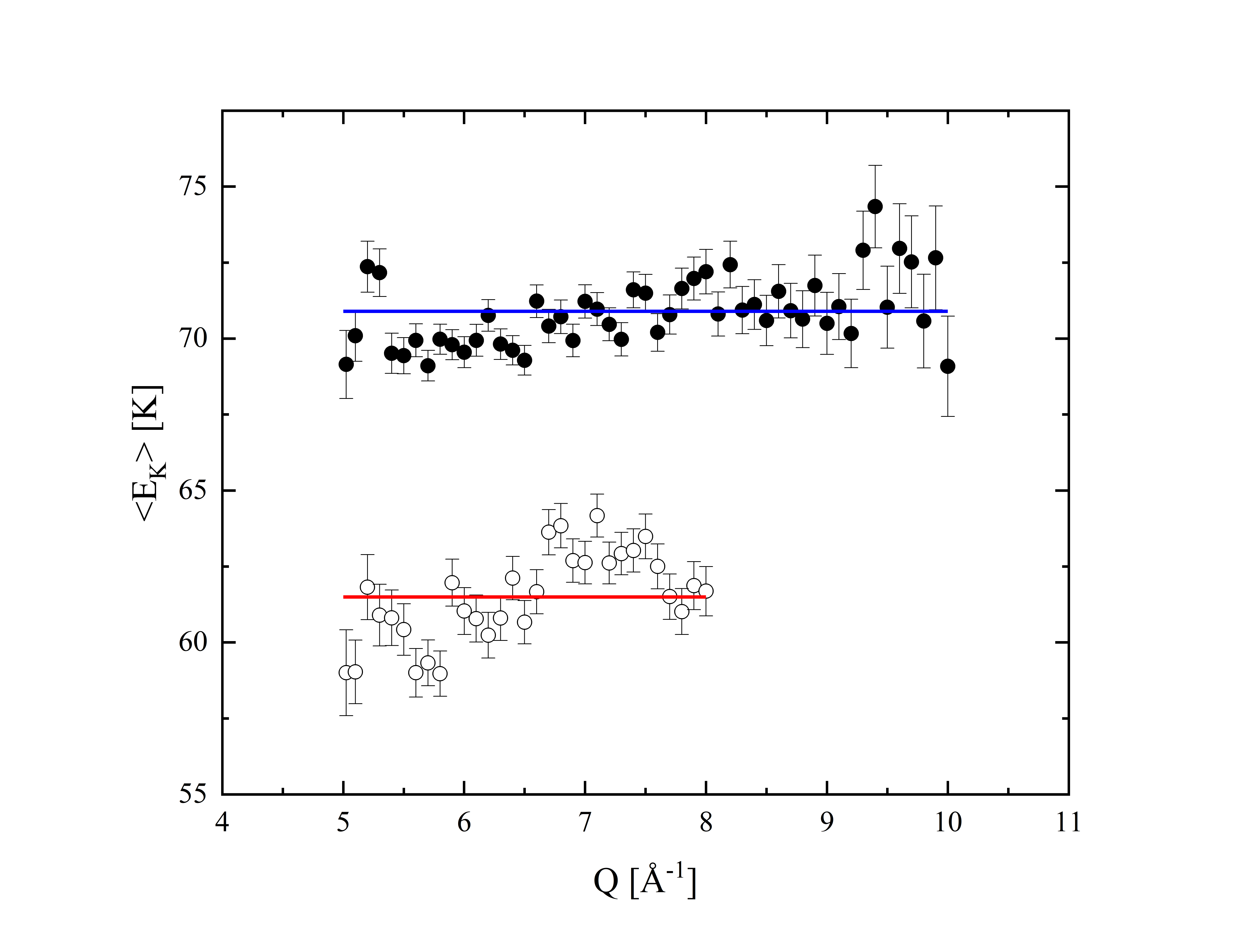}
 \caption{Experimental estimates of $\langle E_K\rangle$ obtained from fitting the scattering to Equation \ref{eq:model} at 5.0 K (closed circles) and 16.5 K (open circles).  The best estimate of $\langle E_K\rangle$ at each temperature is shown by a horizontal line.}
  \label{fgr:KEofQ}
\end{figure}

Figure \ref{fgr:KEofQ} illustrates the kinetic energies extracted from $S(Q, E)$ as a function of $Q$ at two different temperatures.  As expected, the observed kinetic energy is constant with $Q$.  We histogrammed the observed values, adopting the mean as the best estimate of the kinetic energy and the standard deviation as its uncertainty.

\begin{figure}[h]
\centering
  \includegraphics[width=0.9\textwidth,height=0.9\textheight,keepaspectratio]{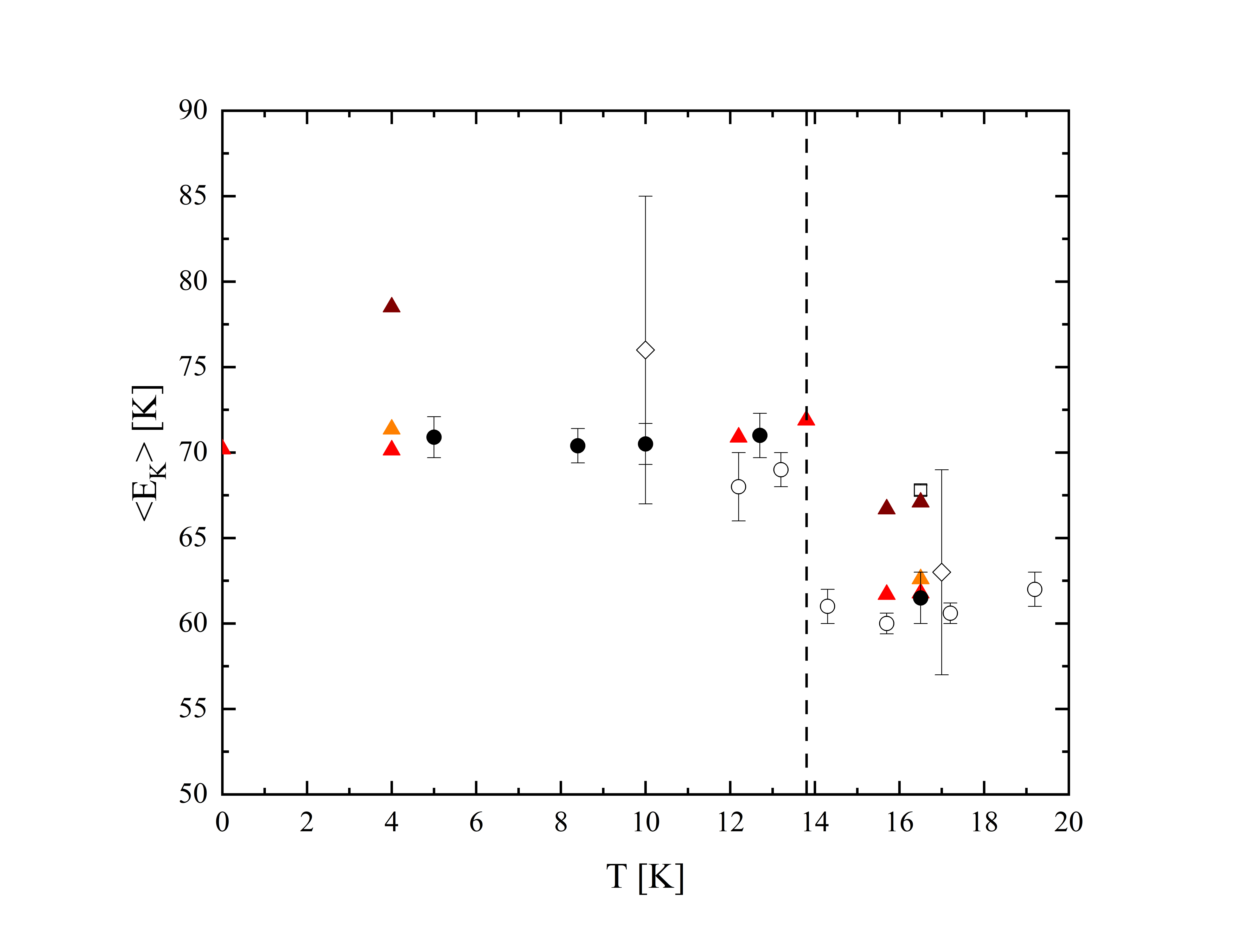}
 \caption{The average molecular kinetic energy of \textit{para}-hydrogen under, or near, saturated vapor pressure.  Experimental estimates: present ARCS study (closed circles); LRMECS\cite{Langel} (open diamonds); TOSCA\cite{CelliKE,ZoppiKE} (open circles); MARI\cite{Dawidowski} (open square).  Quantum Monte Carlo simulations\cite{Boninsegni2009, Dusseault_MSD, H2_potentials} based upon the following model pair potentials: Silvera-Goldman (light red triangles); Buck (orange triangles); and Patkowski (dark red triangles).}
  \label{fgr:KE}
\end{figure}

Our empirical estimates of $\langle E_K\rangle$ are listed in Table \ref{tbl:results} and illustrated in Figure \ref{fgr:KE}.  We find that $\langle E_K\rangle$ is $\approx 70 \textrm{ K}$ from ``low" temperatures up to the triple point.  The average kinetic energy drops to $\approx 62 \textrm{ K}$ in the liquid state.  We contend that this is a manifestation of the Heisenberg indeterminacy principle\cite{Heisenberg}.  At the triple point (13.8 K), solid hydrogen has a molar volume of 23.31 cc/mole whereas the liquid has a molar volume of 26.18 cc/mol\cite{Souers}.  Accordingly, hydrogen molecules are relatively delocalized in the liquid state, and this results in a reduction in the amount of zero-point motion.  This occurs despite the fact that the liquid exists at a higher temperature than the solid.  The reverse effect, namely an increase in average kinetic energy as the density is increased, has been observed in neutron Compton scattering studies of $^3$He\cite{Senesi_3He,Bryan}, and $^4$He\cite{Glyde_P,Diallo2}.

Figure \ref{fgr:KE} compares the present measurements of $\langle E_K\rangle$ with other experimental estimates and with quantum Monte Carlo predictions.  Our empirical estimates of $\langle E_K\rangle$ are in good agreement with the TOSCA findings\cite{CelliKE,ZoppiKE}.  Colognesi \textit{et al} estimate, from the phonon density of states, that $\langle E_K\rangle$ is 68.3(1) K at 13.3 K\cite{H2_DOS}.  In contrast, our results are inconsistent with the outcome of the MARI experiment, from which a kinetic energy of $\approx 68 \textrm{ K}$ at 16.5 K was found.  Also shown in Figure \ref{fgr:KE} are quantum Monte Carlo predictions based upon several different model pair potentials\cite{Boninsegni2009,Dusseault_MSD,H2_potentials}, namely the Silvera-Goldman\cite{SG}, Buck\cite{Buck}, and Patkowski\cite{Patkowski} potentials.  Not shown is the prediction stemming from the Diep-Johnson potential\cite{Diep}, as this model yields an identical prediction to the Patkowski potential at 16.5 K.  There is excellent agreement between the ARCS measurements and simulations based upon the Silvera-Goldman and Buck potentials. 

\subsection{Peak Intensities}
The theory of neutron scattering from molecular hydrogen has been discussed by Sears\cite{Sears1,Sears2} and by Young and Koppel\cite{YK}.  In particular, the latter developed a model for the total neutron scattering cross section of hydrogen, beginning with the assumption that the translational, rotational, and vibrational motions of the molecules are decoupled from one another.   This assumption is motivated by the empirical fact that the rotational states are only weakly perturbed in the liquid and solid states\cite{Silvera,Souers,Raman_pH2}.  Their theory offers an explicit expression for the integrated intensities of the rotational transitions observed in the present experiment:
\begin{equation}
    A^{(YK)}_J(Q) \propto (2J + 1)\lvert a_J(Q) \rvert^2\sigma_J
\end{equation}
\begin{equation}
    a_J(Q) = \int_{-1}^{+1} \exp\left(-\frac{1}{2}\frac{E_R(Q)}{E_{vib}}\mu^2 + iQa\mu\right)    P_J(\mu)d\mu
\end{equation}
Here $\sigma_J$ is the cross section for the $(0, J)$ channel, and it is equal to the coherent (incoherent) scattering cross section of atomic hydrogen when $J$ is even (odd); $E_{vib}$ is the first excited vibrational level; and $P_J$ is the $J^\textrm{th}$ Legendre polynomial.  When $E_R(Q)\ll E_{vib}$, $A^{(YK)}_J(Q)$ reduces to the rotational form factors given above.

The Young-Koppel theory has previously been compared to experiment in the gaseous and condensed phases of hydrogen.  While the theory successfully predicts the integrated intensities $A_J(Q)$ of the gas phase\cite{Herwig_YK}, there remain discrepancies in the condensed phases, at both the level of the peak intensities\cite{Langel}  and the total neutron scattering cross section\cite{H2barns}.  As shown above, the Sears/Young-Koppel form factors correctly describe the integrated intensity of the $(0, 1)$ transition at low energies.  Moreover, in Ref \cite{nH2_diffusion}, it was found that the $(1, 1)$ transition in liquid \textit{normal}-hydrogen exhibits the appropriate Sears/Young-Koppel form factor.

\begin{figure}[h]
\centering
  \includegraphics[width=0.8\textwidth,height=0.8\textheight,keepaspectratio]{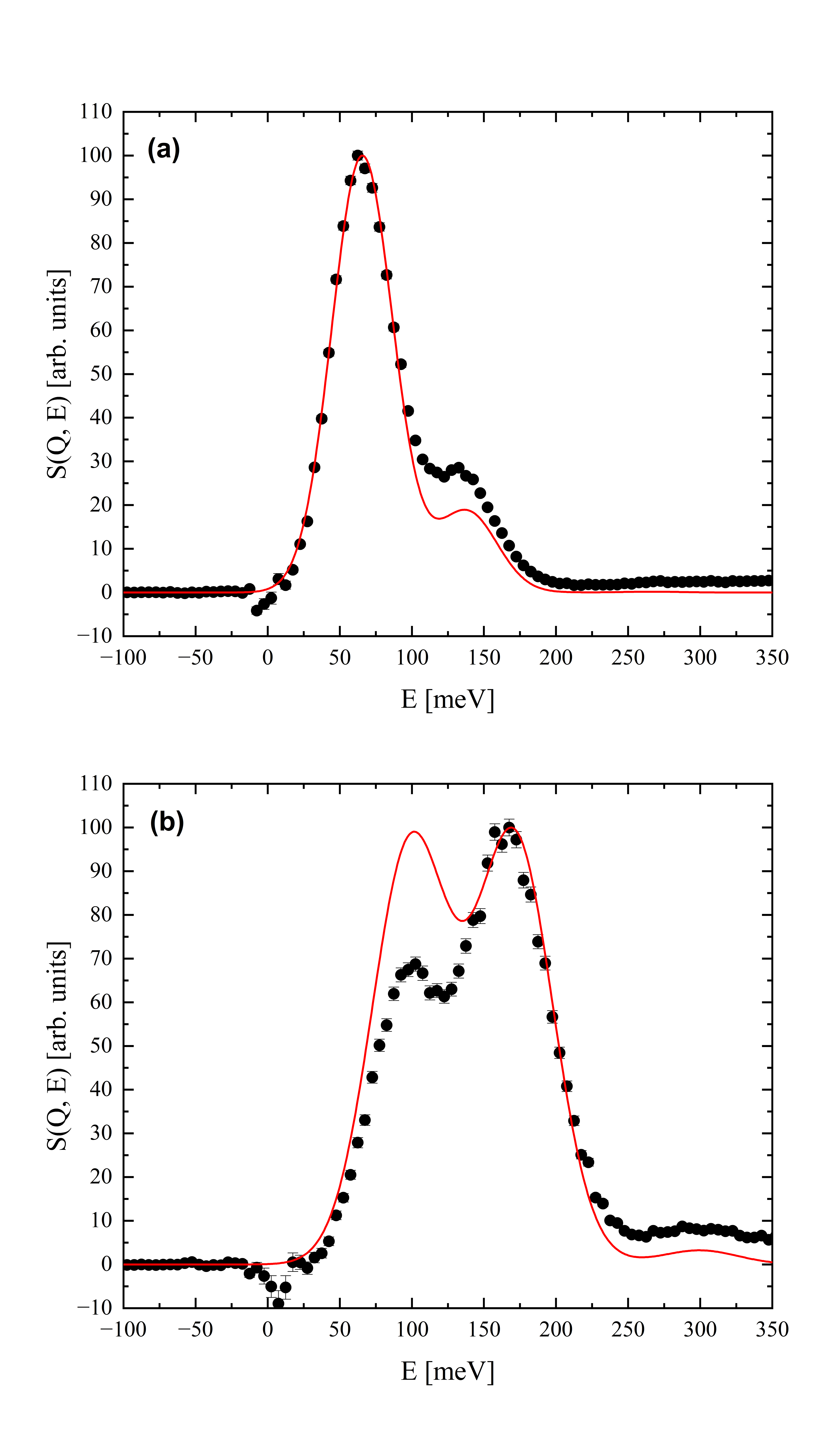}
 \caption{The dynamic structure factor of solid hydrogen at 12.7 K where (a) $Q = 7.0 \textrm{ \AA}^{-1}$ and (b) $Q = 9.0 \textrm{ \AA}^{-1}$.  The solid lines are obtained from the Young-Koppel theory and quantum Monte Carlo predictions for $T = 12.2 \textrm{ K}$, for which $\langle E_K\rangle  = 70.9 \textrm{ K}$.  Both the experimental data and the theoretical curves have been scaled so that the maximum intensity in each panel is 100 arbitrary units.}
  \label{fgr:PIMC_and_YK}
\end{figure}

The validity of the Young-Koppel theory at higher energies will now be considered.  We first combine the Young-Koppel theory with quantum Monte Carlo calculations.  This combination is obtained from Equation \ref{eq:model} by setting the integrated intensities equal to those predicted by the Young-Koppel theory and by setting the intrinsic peak width equal to that predicted by the simulations.  Figure \ref{fgr:PIMC_and_YK} illustrates a representative comparison in the solid phase at two different values of $Q$.  There are no adjustable parameters in this comparison, apart from the intensity scale which has been fixed so that the maximum occurs at 100 units in each panel.  As can be seen, the combined theory is only qualitatively correct: although the peak widths are faithfully reproduced, the peak intensities are not.

\begin{figure}[h]
\centering
  \includegraphics[width=0.8\textwidth,height=0.8\textheight,keepaspectratio]{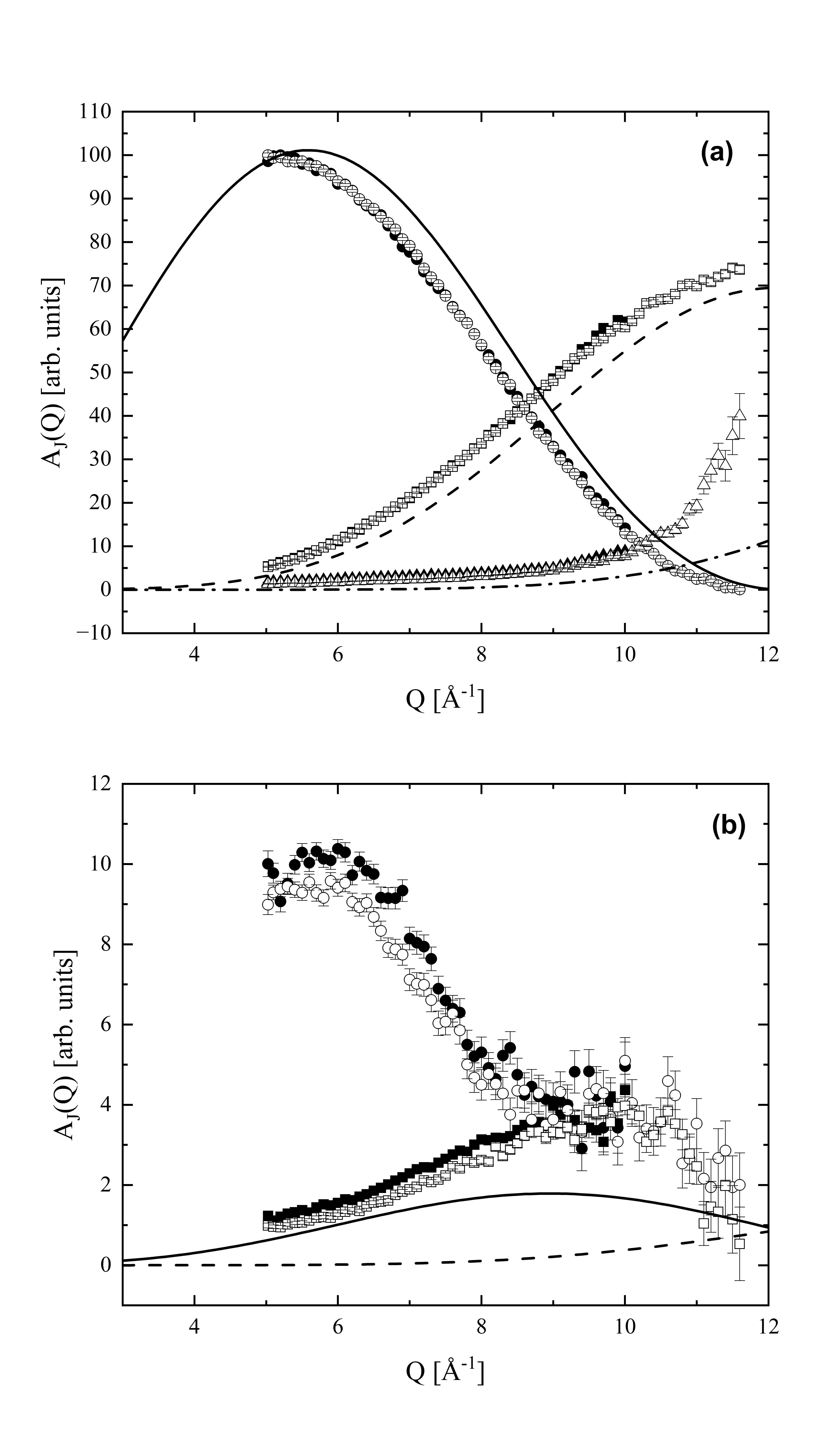}
 \caption{The integrated intensities $A_J(Q)$ estimated by fitting Equation \ref{eq:model} to the 5.0 K data set.  One obtains the closed symbols when the kinetic energy is allowed to be an adjustable parameter, whereas one obtains the open symbols when the kinetic energy is fixed to the quantum Monte Carlo value.  In panel (a), transitions to $J = 1, 3, 5$ are shown as circles, squares, and triangles, respectively.  In panel (b), transitions to $J = 2, 4$ are shown as circles and squares, respectively.  In both panels, the continuous lines are predictions from the Young-Koppel theory.}
  \label{fgr:AofQ}
\end{figure}

In Figure \ref{fgr:AofQ}, we return to the outcome of the fits to Equation \ref{eq:model} where the integrated intensities are treated as free parameters.  Panel (a) shows that the Young-Koppel theory is in semi-quantitative agreement with the observed intensities for transitions to odd-$J$ states.  The deviations are consistent with those first reported by Langel \textit{et al}\cite{Langel}.  Panel (b) compares the predictions of the Young-Koppel theory with the observed intensities for transitions to the even-$J$ states.  Here there are more striking, and perhaps more surprising, differences between theory and experiment.  In previous studies\cite{Langel,CelliKE,ZoppiKE,Dawidowski}, transitions to even-$J$ were not considered in the data analysis, either because they were thought to have negligible intensity or because the available energy resolution was too coarse to observe them.  The present study was carried out with an energy resolution three times sharper than that of Ref \cite{Langel}, apparently allowing for the contribution of the even-$J$ states to the neutron scattering spectrum to be observed.

\section{Conclusions}
In this paper, we presented an inelastic neutron scattering study of liquid and solid hydrogen under saturated vapor pressure.  We obtained high-precision empirical estimates of the molecular mean-squared displacement and the average translational kinetic energy.  Both quantities are largely shaped by quantum-mechanical zero-point motion.  In the solid state, the mean-squared displacement increases from $0.479(5) \textrm{ \AA}^2$ at 5.0 K to $0.536(5) \textrm{ \AA}^2$ at 12.7 K, an increase of $\approx12\%$.  Across the same temperature range, the average kinetic energy of the hydrogen molecules is, to within current precision, constant.  It drops precipitously in the liquid state, going from $71.0 \pm 1.3$ K at 12.7 K to $61.5 \pm 1.5$ K at 16.5 K.  The reduction in $\langle E_K\rangle$ is a consequence of the indeterminacy principle together with the large increase in the molar volume of the substance upon melting.

The results of the present study may be compared with both theoretical predictions and with previous experiments.  In general, we find good agreement between our measurements of the molecular mean-squared displacement and average kinetic energy with quantum Monte Carlo simulations based upon the Silvera-Goldman and Buck potentials.  Simulations proceeding from the Patkowski potential overestimate the average kinetic energy in both the liquid and solid states by $\approx10\%$.  Our results provide independent confirmation of the empirical estimates of the kinetic energy obtained from TOSCA, whereas they offer disconfirmation of the IN20 and MARI studies. 

In our view, the outcome of this experiment sheds new light upon the liquid-solid phase transition of molecular hydrogen, and upon the reliability of numerical models of condensed hydrogen built upon currently available model potentials.

\section*{Acknowledgements}
The authors thank Craig Brown for scientific discussion about this project, and Luke Daemon for assistance with the hydrogen gas handling system.  A portion of this research used resources at the Spallation Neutron Source, a DOE Office of Science User Facility operated by the Oak Ridge National Laboratory.  Certain commercial equipment, instruments, or materials are identified in this paper to foster understanding. Such identification does not imply recommendation or endorsement by the National Institute of Standards and Technology, nor does it imply that the materials or equipment identified are necessarily the best available for the purpose.

\bibliography{hydrogen}

\newpage

\section*{Figures and Tables}

\begin{table}
\caption{  \label{tbl:results} Summary of present results.}
\begin{ruledtabular}
\begin{tabular}{cccccc}
$T$ [K] & $n$ [$\textrm{\AA}^{-3}$] & $\langle u^2\rangle$ [$\textrm{ \AA}^{2}$] & $\alpha$ [$\times 10^{-3} \textrm{ \AA}^{2}$] & $\delta$ &  $\langle E_K\rangle$ [K]\\
\hline
    5.0 & 0.0261 & 0.479(5) & 1.09(10) & 0.183 & $70.9 \pm 1.2$  \\
    8.4 & 0.0261 & 0.495(5) & 1.38(13) & 0.186 & $70.4 \pm 1.0$  \\
    10.0 & 0.0260 & 0.505(5) & 1.47(10) & 0.188 & $70.5 \pm 1.2$  \\
    12.7 & 0.0259 & 0.536(5) & 1.68(11) & 0.193 & $71.0 \pm 1.3$  \\
    16.5 & 0.0223 & -- & -- & -- & $61.5 \pm 1.5$  \\
\end{tabular}
\end{ruledtabular}
\end{table}

\end{document}